# The Scalable Brain Atlas: instant web-based access to public brain atlases and related content.

**Rembrandt Bakker (1,2), Paul Tiesinga (1), Rolf Kötter*(3)**

*Rolf Kötter sadly passed away on June 9th, 2010. He co-initiated this project and played a crucial role in the design and quality assurance of the Scalable Brain Atlas

1. Department of Neuroinformatics, Donders Institute for Brain, Cognition and Behavior, Radboud University Nijmegen, Nijmegen, Netherlands

2. Institute of Neuroscience and Medicine (INM-6) and Institute for Advanced Simulation (IAS-6), Jülich Research Centre and JARA, Jülich, Germany

3. Centre for Neuroscience, Donders Institute for Brain, Cognition and Behavior, Radboud University Nijmegen Medical Centre, Nijmegen, The Netherlands

## Abstract

The Scalable Brain Atlas (SBA) is a collection of web services that provide unified access to a large collection of brain atlas templates for different species. Its main component is an atlas viewer that displays brain atlas data as a stack of slices in which stereotaxic coordinates and brain regions can be selected. These are subsequently used to launch web queries to resources that require coordinates or region names as input. It supports plugins which run inside the viewer and respond when a new slice, coordinate or region is selected. It contains 20 atlas templates in six species, and plugins to compute coordinate transformations, display anatomical connectivity and fiducial points, and retrieve properties, descriptions, definitions and 3d reconstructions of brain regions. The ambition of SBA is to provide a unified representation of all publicly available brain atlases directly in the web browser, while remaining a responsive and light weight resource that specializes in atlas comparisons, searches, coordinate transformations and interactive displays.

## 1. Introduction

Brain atlases are used in all areas of neuroscience, and an enormous amount of research data that is tied to coordinates in the brain is produced every day in laboratories worldwide. Many initiatives exist to make these data available through public databases. Federated access to these resources is provided by the Neuroscience Information Framework (NIF, Gardner et al. 2008). The NIF provides services for structured, ontology-based queries, but these are impractical for accessing spatially registered content. For such data, a brain atlasing framework is needed that allows 1) spatial navigation through the brain to select a region or coordinate to initiate a database query, and 2) display of returned results in stereotactic space. Prime examples of existing solutions for these tasks are the Brain Explorer software from the Allen Institute (Sunkin et al. 2013), the (discontinued) Brain Navigator product of Elsevier Inc. (http:// brainnav.com), the Java-based Whole Brain Catalog (http://wholebraincatalog.org), the urfacebased analysis package Caret (Van Essen 2011), the JuBrain Cytoarchitectonic Atlas Viewer Mohlberg et al. 2012), the McGill BrainBrowser (https://brainbrowser.cbrain.mcgill.ca/), the NeuroMaps atlas viewer and registration tool (Dubach and Bowden 2009), the Mouse BIRN Atlasing toolkit (Lee et al. 2010), the Three-Dimensional Rodent Atlas System (Hjornevik et al. 2007), and the neuroVIISAS integration and simulation platform (Schmitt and Eipert 2012). In principle, each of these products can retrieve and display spatial brain data. What is lacking however is a platform that is 1) not tied to a particular atlas, vendor, database or species, 2) runs in the web browser without having to install software, and 3) allows bilateral interaction with online data resources. The Scalable Brain Atlas (SBA) addresses these issues by using open web standards and having the ambition to contain all publicly available brain atlases that are of sufficient interest to the community.

### 1.1 Web-based interactive brain atlas

The SBA has evolved as the successor of the CoCoMac-Paxinos-3d tool (CP3D, Bezgin et al. 2009), which is a Java-based platform that volume-renders brain regions taken from the Paxinos rhesus monkey atlas (Paxinos et al. 2000) and displays structural connectivity data from the CoCoMac database (Stephan et al. 2001, Kötter 2004) as directed arrows. While converting CP3D to a fully web-based service, we decided to simplify its 3d requirements because support for 3d rendering in web browsers is still in its infancy and the required bandwidth restricts its applicability. We instead use a quasi 3d approach, whereby sets of 2d drawings are stacked together to create a 3d experience. Several technologies exist to interactively render such complex drawings inside a web browser, such as Adobe Flash (Adobe Systems Inc.), Microsoft Silverlight (Microsoft Corporation), and Scalable Vector Graphics (SVG, Dahlström et al. 2011). We selected SVG for the SBA because it is an open standard and has broad cross browser support.

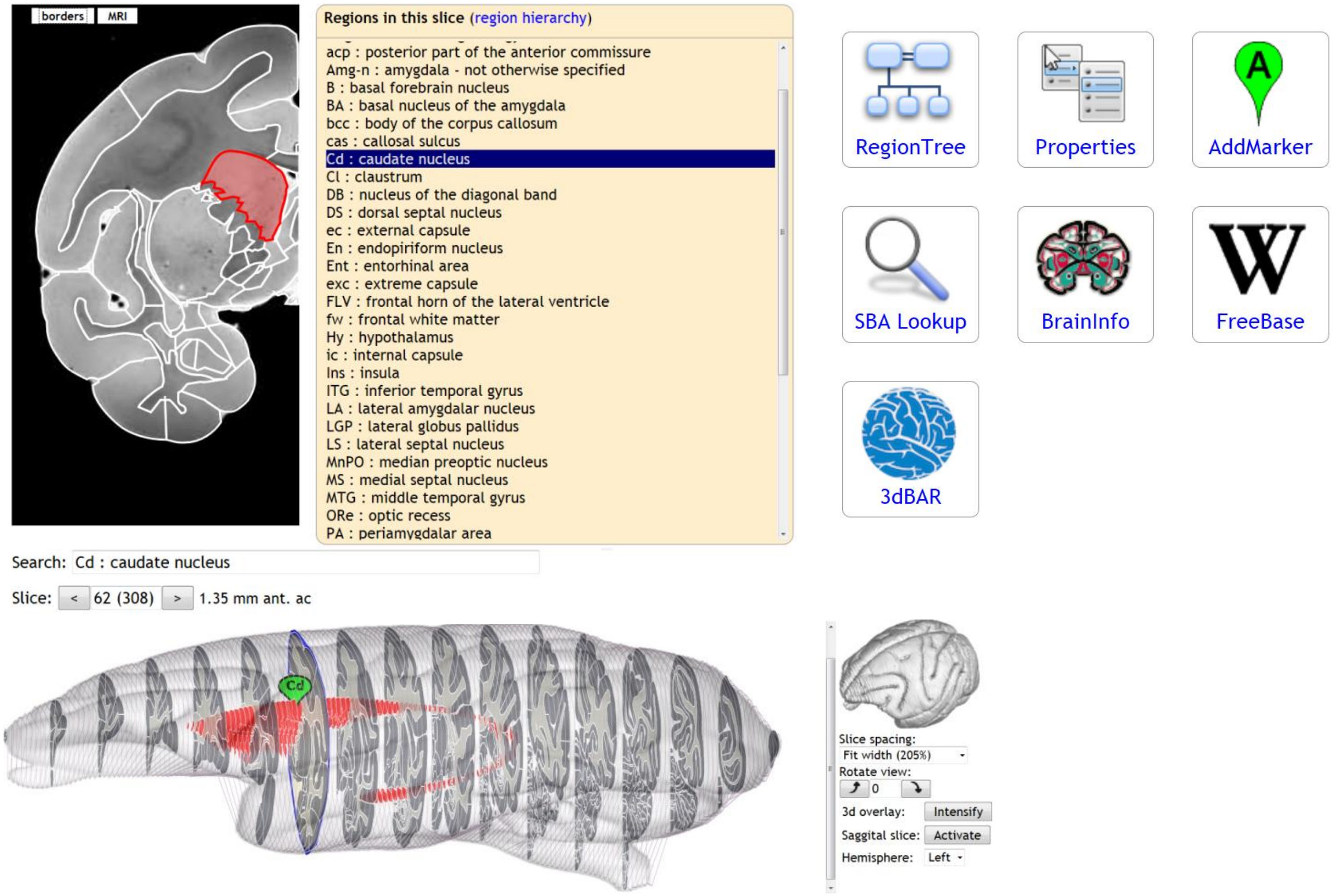

**Figure 1. Components of the atlas viewer: (A) 2d panel displaying a single slice with the parcellation overlaid on the selected imaging modalities, along with a list of structures in the current view or the full region hierarchy (as in Fig. 3); (B) 3d panel with convex hulls for each slice, and the detailed parcellation for every 10th slice. The current slice is highlighted as a blue contour. By default, the brain is elongated to better show the inside. The 3d surface rendering overlay is a static image, generated by the 3dBAR service (Sec. 5.1), with adjustable transparency; The marker named 'Cd' is created by the AddMarker plugin; (C) Plugin panel. When a plugin gets activated, it responds to changes in selected region, slice and coordinate.**

The CP3D tool was tied to a particular species (Macaque) and application (CoCoMac). With the creation of its SVG-based counterpart, we generalized the tool and renamed it to Scalable Brain Atlas. It is scalable because (1) it supports multiple species and multiple brain atlases per species; (2) it has a plugin architecture that allows bidirectional interaction with web-based resources; (3) it is based on SVG. At present, twelve plugins are operational and twenty different brain atlases have been imported. Atlas providers are encouraged to submit data for inclusion. The SBA is hosted at [http://scalablebrainatlas.incf.org](http://scalablebrainatlas.incf.org).

## 1.2 Core features of the SBA

At its core, the SBA has all the features of a typical paper version of a brain atlas, with a 2d view of a selected slice and its delineated brain structures (Fig. 1A). It also has a 3d panel which contains a stack of all slices and shows the full extent of a selected region (Fig 1B). A wide variety of mouse actions and keyboard controls are available to navigate, search and display brain regions. The SBA shows the stereotaxic coordinates of the mouse pointer, and markers can be attached to selected locations.

Clicking on a slice in the 3d panel opens it in the 2d panel. The 2d panel shows the brain region delineations in a single slice, which can be underlaid with one of the available imaging modalities. Clicking anywhere in the 2d panel triggers a range of actions: (1) The region and all its subparts get highlighted in the 2d panel; (2) The region and all its subparts in all slices where the region appears get highlighted in the 3d panel. (3) The region gets highlighted in the (hierarchical) list of regions for the given atlas. (4) The active plugin receives a trigger, and can use either the newly selected region or the stereotactic coordinate of the mouse click to update its contents. The 3d panel has controls to rotate the view, stretch it in the slice dimension, and to overlay the slice stack with a pre-rendered 3d surface representation of the brain (see 3dBAR plugin). In addition to this, a framework for displaying stereotaxic markers in the 3d panel is available for use by the various plugins.

Ten plugins are currently included in the SBA core. They are arranged as tiles next to the slice panel (Fig 1C), and provide functionality such as displaying connectivity derived from the CoCoMac database, getting 3d surface renderings of selected brain regions, and showing locations of anatomical landmarks. New plugins can be developed and tested after one-time registration at the SBA server.

In addition to  the plugins that interact directly with the graphical interface, the SBA provides web services that allow  other websites to retrieve atlas-derived data and images. The most important services are (1) conversion of SVG-based renderings to bitmap images; (2) export of atlas delineations to label volumes that can be analyzed with Matlab; (3) generating a hierarchical list of regions for each imported atlas; (4) computation of several metrics such as region centers and distance matrix. At the most basic level, all available atlas templates are accessible as downloadable files, whose structure is outlined in Appendix A.

The core components of the SBA are invisible to the user and include routines to import new sets of atlas data into the system. The SBA works with coronal slices and for each slice the outline of each delineated region needs to be provided as a closed curve. Many atlas sources come in the form of labelled volumes (where the voxel color represents the region name), from which the curves need to be traced.

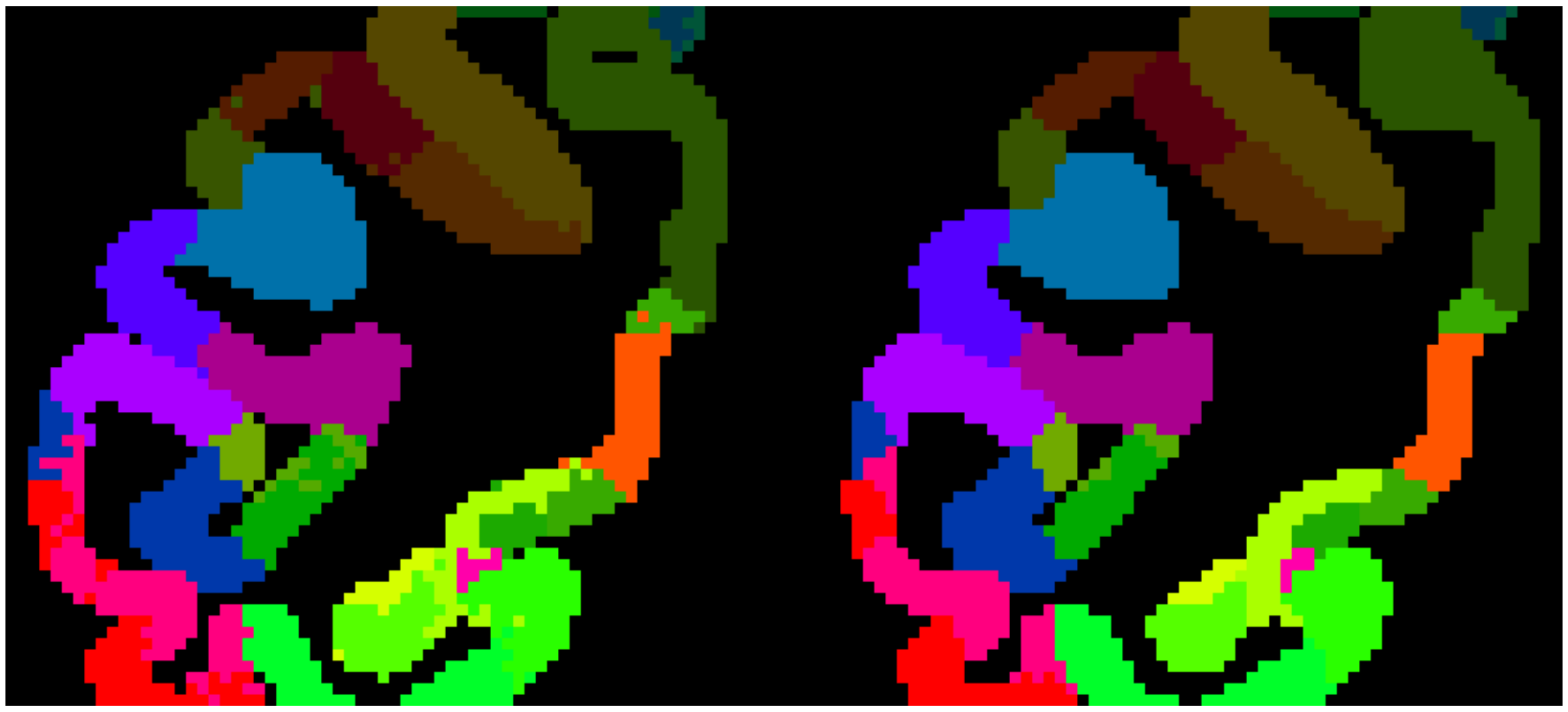

**Figure 2. Effect of the 3d smoothing kernel that is applied to volumes obtained after Surface to Volume conversion in Caret (Van Essen, 2011). The blurring kernel is applied separately in each dimension with coefficients [1, 3, 6, 3, 1]. It effectively despeckles the 3d volume.**

In the following section we discuss the methods used to create atlas templates, interactive web pages, and plugins. We then present an overview of the plugins and services that are currently available at http://scalablebrainatlas.incf.com. We thereby emphasize how each of them contributes to the goal of the SBA to query online resources that contain brain-region or brain-coordinate related content. In the discussion we highlight the strength of the SBA as a web-based data display engine, and outline further work that would facilitate the data integration across atlas templates, across databases, and across species.

# 2. Methods

## 2.1 Importing atlases into the SBA

The SBA currently contains twenty atlases, listed in Table 1. One of these (PHT00) was available in SVG format from the CP3D project, wherein the SVG polygons were created by manual tracing of scanned atlas pages. For paper atlases that have their document source available, the polygons can be derived automatically if the source is vector-based. Two issues emerge when parsing the file automatically: (1) lines in vector drawings often do not form closed regions; and (2) labels are not always placed inside regions due to space constraints. These issues were largely addressed by the smart parsing methods of Majka et al. (2012), who also proposed a Common Atlas Format (CAF) to store the processing results and related metadata. We created a filter to import CAF files into the SBA, and used this to import the Marmoset atlas (PWPRT12). Many other high quality print atlases could potentially be imported in this way, but in most cases the copyright has been transferred to the publisher, blocking a public release.

Most of the imported atlases are obtained from label volumes, wherein the color index of each voxel represents the region that it belongs to. Such volumes are typically stored in the NIfTI format (Cox et al. 2004), which has the benefit that the scale and origin of the brain space are included. NIfTI volumes were converted to stacks of coronal bitmap images using the Matlab NIfTI toolbox

(http://research.baycrest.org/~jimmy/NIfTI/). We tested several off-the-shelve tools to trace the contours of color coded regions. The open source software potrace (http://potrace.sourceforge.net) is easily integrated into processing pipelines, but the borders of adjacent regions are individually parameterized, which causes small gaps or overlap. These issues are solved by using the 'PowerTrace' routines of CorelDraw X4 (Corel Corporation). To prevent PowerTrace from merging regions with similar colors, adjacent regions must be assigned highly contrasting colors.

Six imported atlases are derived from cortical surface parcellations downloaded from the SumsDB brain-mapping repository (Van Essen, 2002), processed by the software package Caret (Van Essen, 2011). The first step is to convert the labelled surfaces to a label volume, by discretizing the space and assuming a constant cortical thickness; we used 1.8 mm for Macaque data. In the resulting label volumes, the region boundaries appear 'ragged' (Fig. 2A). We smoothed the regions by first decomposing the volume into individual regions, then applying a blurring filter to each of them, and then regenerating the volume by choosing for each voxel the region with the highest intensity. Figure 2B illustrates the smoothing effect of this procedure. The smoothed label volumes were further processed as in the previous paragraph.

## 2.2 Creating interactive SVG-based webpages

Web browsers have a long tradition in displaying structured text documents, formatted according to the Hypertext Markup Language (HTML) specification (Raggett and Le Hors 1999). This specification deals with text, bitmap images, layout, and hyperlinks. XML (Bray et al. 2008) is the generic container format for languages such as HTML, and SVG is an XML-based specification for vector graphics. It is supported by all major web browsers.

In an interactive web page, content dynamically responds to keyboard and mouse controls. The response can be: (1) fully client-side, and involve only page elements that are already loaded; or (2) a client-server interaction: retrieve new content from a server and display the result in the client. The SBA plugins use client-server interaction, but in the SBA-core all interactivity is client side. This has the advantage that no internet connection is required once the page is loaded; the downside is that the page may take a while to load; the atlas templates in the SBA are 1 to 2MB in size, most of which is taken up by the polygons that define region shapes. The server contains a caching mechanism that stores gzip-compressed atlas pages to reduce page load time by about two thirds.

JavaScript (ECMA-262 2011) is the dominant technology to drive client side interactivity. When opening an atlas, the complete set of Javascript Object Notation (JSON) files that define the atlas template (Appendix A) are downloaded at once and stored in JavaScript memory. JavaScript code then generates a mixture of HTML and SVG, and renders the page in the browser. Inside JavaScript, the page is stored as a hierarchical tree known as the Document Object Model (DOM), which covers both the HTML and SVG elements. Following a mouse click or key press, JavaScript code changes relevant parts of the DOM, which is directly reflected on the displayed page.

Client-server interaction happens when the page is first loaded, and whenever a plugin initiates a web query. Pages typically have a URL that contains an address plus a query that encodes a list of parameters. Many different server-side technologies are available to dynamically respond to these parameters (e.g., the selected atlas template). The SBA uses the PHP scripting language (http://php.net), running on an Apache web server (http://httpd.apache.org/). When opening an atlas, PHP code generates a HTML page that contains a mixture of scripts (JavaScript), data for the selected atlas (JSON), markup (HTML) and style.

## 2.3 Creating plugins

The SBA has part of its browser window reserved for plugins, which are small applications that are triggered when the user selects a new region, slice, or stereotaxic coordinate.  Following the trigger, the plugin can change the content of its own frame, or it can call routines from the SBA viewer library to change the content of the 2d and 3d panels of the SBA. For example, it can create a marker using 'new sbaMarker_class(...)', and display it at a stereotaxic location or region center in both the 2d and 3d panel. An example plugin that prints the text 'HELLO WORLD' is listed in Appendix B, which also provides pointers to the source code of the existing plugins. Writing a plugin does require an understanding of Javascript prototypes.

If no plugin is activated, the plugin window presents a list of all available plugins for the given atlas template. To become part of that list, a plugin has to be approved and hosted at the SBA website.

### Bilateral client-server communication

When a plugin responds to a state change of the atlas viewer, it can use the Javascript httpRequest method to send requests for downloading content. For browser security reasons, Javascript has a same origin policy (SOP), allowing only requests to the server that hosts the website. If a plugin needs to access content from external sites, the solution is to define a PHP script on the SBA server that passes on the request to the external site, and returns the result to the client. This is how it is implemented for the CoCoMac, NeuroLex, Wikipedia and DAI plugins (Sec. 5). Note that the SOP can be bypassed with JSONP (https://en.wikipedia.org/wiki/JSONP), a protocol that disguises data as code, which is exempt from the SOP.

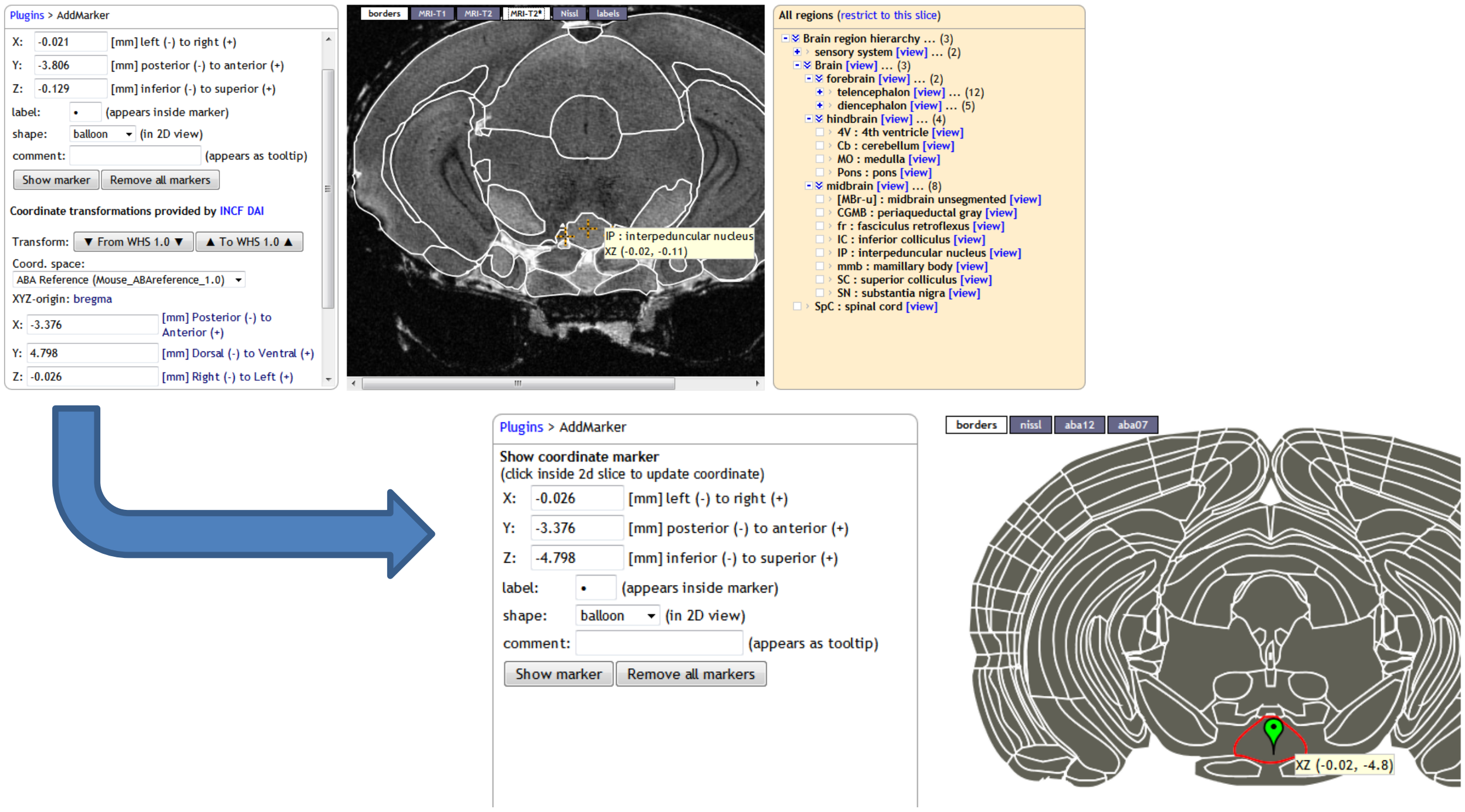

**Figure 3. Coordinate transformation invoked from the AddLandmark plugin, powered by the INCF Digital Atlasing infrastructure. Here, the interpeduncular nucleus landmark is transformed from the WHS12 to the ABA12 template. Table 1 explains template names.**

The plugins themselves do not need to be hosted on the SBA server, because the SOP does not apply to Javascript files. Externally hosted plugins can be imported by adding its full URL as a query parameter, for example *plugin=http://www.mysite.org/myplugin.js*. External plugins must be white-listed on the SBA server to prevent insertion of malicious code.

### 2.4 Creating services

Services are scripts designed to serve content to other websites, typically called by a URL and a set of query parameters. The SBA uses a self-documenting service framework: If the service is called with missing parameters, a form is presented with the names and admissible values for these parameters. Each service contains a header section that is used by the sitemap.php service to generate an annotated list of all available services.

## 3. Imported atlas templates

An important goal of the SBA is to provide unified access to publicly available atlas templates. To be included, a template must meet four or more of the following criteria:

- is publicly accessible
- is described in a peer-reviewed publication
- contains both a brain parcellation and underlying data modality
- is part of a resource that contains valuable neuroscience data
- is (becoming) a standard reference space
- is available in parseable format (NIfTI-1, CAF)

Table 1 lists the twenty atlas templates that have so far been included, covering six species. The original goal of being a CoCoMac connectivity viewer has caused the Macaque to be overrepresented. Underrepresented are atlases for which copyrights have been transferred to publishers.

### 3.1 Citation policy

The SBA processes and integrates atlas templates from many different publicly available sources. If researchers prefer the SBA-processed templates over the original sources, they might be tempted to cite the SBA as the source of an atlas template. To protect the scientific careers of those who created the atlas, the SBA requires its users to always cite the 'defining publications' written by the creators of the template, even when atlases are transformed or combined.

### 3.2 Waxholm Space

Of special interest are the Waxholm Space templates WHS12 (C57BL/6 mouse) and PLCJB14 (Sprague Dawley rat), promoted by the International Neuroinformatics Coordinating Facility (INCF) as standard reference spaces, and defined by high resolution (21.5 µm isotropic for WHS12) MR imaging data. The purpose of having a standard reference space is to make it easier to transfer data between atlases: if two atlases both have a mapping to the standard space defined, then 1) the mapping between the two atlases is implicitly defined, and data available in each atlases can be displayed together in the standard space. A future role for the SBA can be to recommend standard templates, and to compute optimal (non)linear transforms to and from these standard templates. A preview of this feature is shown in Fig 3, where the *interpeduncular nucleus middle* landmark is transformed from the WHS12 to the ABA12 space. The INCF Digital Atlasing Infrastructure (DAI, Hawrylycz et al. 2011) is queried to obtain the transformation between the two templates.

## Table 1: Available atlas templates

See http://scalablebrainatlas.incf.org/services/listtemplates.php for the most current list, including templates under development.

| *template* | *title* | *Primary publication/site* | *Parcellation* | *Imaging modalities* |
|---|---|---|---|---|
| **Mouse** | | | | |
| ABA2012 | Allen Mouse Brain 2012 | Dong (2008) | 667 areas incl. layer subdivision | Nissl |
| WHS12 | Waxholm Space atlas 2012 | Johnson et al. (2010) | 39 areas | Nissl and 21.5 µm resolution MR (T1, T2w, T2*) |
| | | | | |
| **Rat** | | | | |
| PLCJB14 | Waxholm space Sprague Dawley reference atlas | Papp et al. (2014) | 97 areas (neocortex=1 area) | 39 µm T2*, DTI, DWI, fractional anisotropy |
| CBWJ13_age_P80 | MR-Histology atlas at postnatal day 80 | Calabrese et al. (2013) | 27 areas | 25 µm resolution MR (T2*/GRE) |
| VSNetal11 | Wistar rat in vivo MRI template | Valdés-Hernández et al. (2011) | 129 cortical areas | T2w, white/gray matter, csf |
| RMJetal13_age_P72 | DTI Atlas of the Rat Brain (age P72) | Rumple et al. (2013) | 29 areas | 160 µm DTI |
| VLAetal11 | Population-averaged DTI atlas | Veraart et al. (2011) | 14 areas | T1w, DWI, FA |
| | | | | |
| **Marmoset** | | | | |
| PWPRT12 | Marmoset Cortical structures provided by M. Rosa | Paxinos et al. (2012) | 116 cortical areas | Nissl, plus seven other stains via marmoset-brain.org |
| **Macaque** | | | | |
| PHT00 | Rhesis monkey in stereotaxic coordinates | Paxinos, Huang, Toga (2000) | 283 areas[a]: cortex, amygdala, thalamus, striatum | - |
| DB08 | NeuroMaps Macaque atlas | Dubach, Bowden (2009) | 384 anatomically defined areas | T1 |
| FVE91_on_F99[b] | Felleman and VanEssen 1991 in F99 space | Felleman, Van Essen (1991) | 73 cortical areas | T1 |
| LVE00_on_F99[b] | Lewis and VanEssen 2000 in F99 space | Lewis, Van Essen (2000) | 87 cortical areas | T1 |
| MMFetal11_on_F99[b] | Markov et al. 2011 in F99 space | Markov et al. (2011) | 81 cortical areas | T1 |
| MERetal12_on_F99[b] | Markov et al. 2012 in F99 space | Markov et al. (2012) | 93 cortical areas | T1 |
| RM_on_F99[b] | Regional Map in F99 space | Kötter and Wanke (2005) | 41 anatomical areas | T1 |
| | | | | |
| **Opossum** | | | | |
| OPSM14 | Multimodal atlas of gray short-tailed opossum brain | Majka et al. (2013); Chlodzinska et al. (2013) | 105 areas (neocortex=1 area) | |
| | | | | |
| **Human** | | | | |
| EAZ05 | JuBrain cytoarchitectonic parcellation | Eickhoff et al. (2005) | 76 cyto-architectonic areas | averaged MRI template |
| LPBA40_on_SRI24 | LBPA40 areas in SRI24 space | SRI24: Rolfing et al. (2010) LBPA40: Shattuck et al. (2008) | 56 cortical areas incl. Left/Right division | T1w, T2w, rho |
| B05_on_Conte69 | Brodmann[d] areas in Conte69 space | Glasser, Van Essen (2011) | 47 Brodmann cortical areas | T1w, T2w, T1w/T2w |
| BIGB13 | Bigbrain, resampled at 400 µm | Amunts et al (2013) | - | Nissl, resampled at 400 µm |

[a] This is a subset of the print atlas, which contains many more subcortical structures.

[b] Obtained as a cortical surface from the SumsDB repository (http://sumsdb.wustl.edu), and converted to volumetric data by assuming a constant cortical thickness of 1.8mm, using Caret software (Van Essen 2012).

[c] The F99 space is based on a 0.5 mm resolution MR scan (Van Essen 2002).
[d] Brodmann areas refer to the cytoarchitectonic brain parcellation by Brodmann (1909).
Abbreviations: MRI – Magnetic Resonance Imaging; T1, T1w, T2w, T2*, and rho are MRI contrasts that are sensitive to different tissue properties; GRE – Gradient Echo sequence; FA – Fractional Anisotropy.

# 4. Services

SBA services are invoked as URL queries, and return formatted content to SBA plugins, clients (end users), or other websites. They typically perform an operation on atlasing data within the SBA, and return the result as an image, JSON data, or web page.

The complete list of services is available from the sitemap http://scalablebrainatlas.incf.org/sitemap.php. We here describe six core services, their key parameters and intended use. Services are called as http://scalablebrainatlas.incf.org/folder/servicename.php?param1=value1&param2=value2 etc. Documentation is displayed when calling a service without parameters.

## 4.1 Atlas viewer (main/coronal3d.php)

This is the main interactive atlas viewer as described in Sec. 1.2 and displayed in Fig. 1. It has one required parameter *template*, specifying the atlas template to load. Other useful parameters are:

- region (the brain region to highlight). It is first matched with the list of region abbreviations that comes with the template. If that fails, the alias list is search, then the full names, and finally a case-insensitive match is tried.
- plugin (the plugin to activate). If the plugin starts with http:// or https://, it is assumed that an external plugin is intended. External plugins must be white listed to prevent abuse.
- underlay2d (the image modality to display in the slice panel).

## 4.2 List atlas templates (services/listtemplates.php)

Returns an html table with all available atlas templates, the species that they apply to, and the atlas space that the template is registered to. If the atlas space is *native*, the template defines its own space.

## 4.3 List atlas regions (services/listregions.php)

Returns a tab-separated table with all regions defined for the given *template*. The table includes full names, parent acronyms, and a shape code that specifies whether the region is *visible* in the atlas viewer.

## 4.4 Coordinate to region (services/coord2region.php)

Returns the region name that matches the specified stereotactic coordinate, for the given *template*. The *coord* parameter specifies a comma separated triplet x, y and z. Their origin and direction depends on the template, but we adhere to the NIfTI-1 standard in that *x*, *y* and *z* represent the left/right, posterior/anterior and inferior/superior axis, respectively. The service uses ImageMagick (http://www.imagemagick.org) to convert the coronal slice that corresponds to the *y* parameter to a raster image, looks up the color value of the pixel that corresponds to the *x,z* location, and finds the corresponding region name in the *rgb2acr.json* list (see Appendix A).

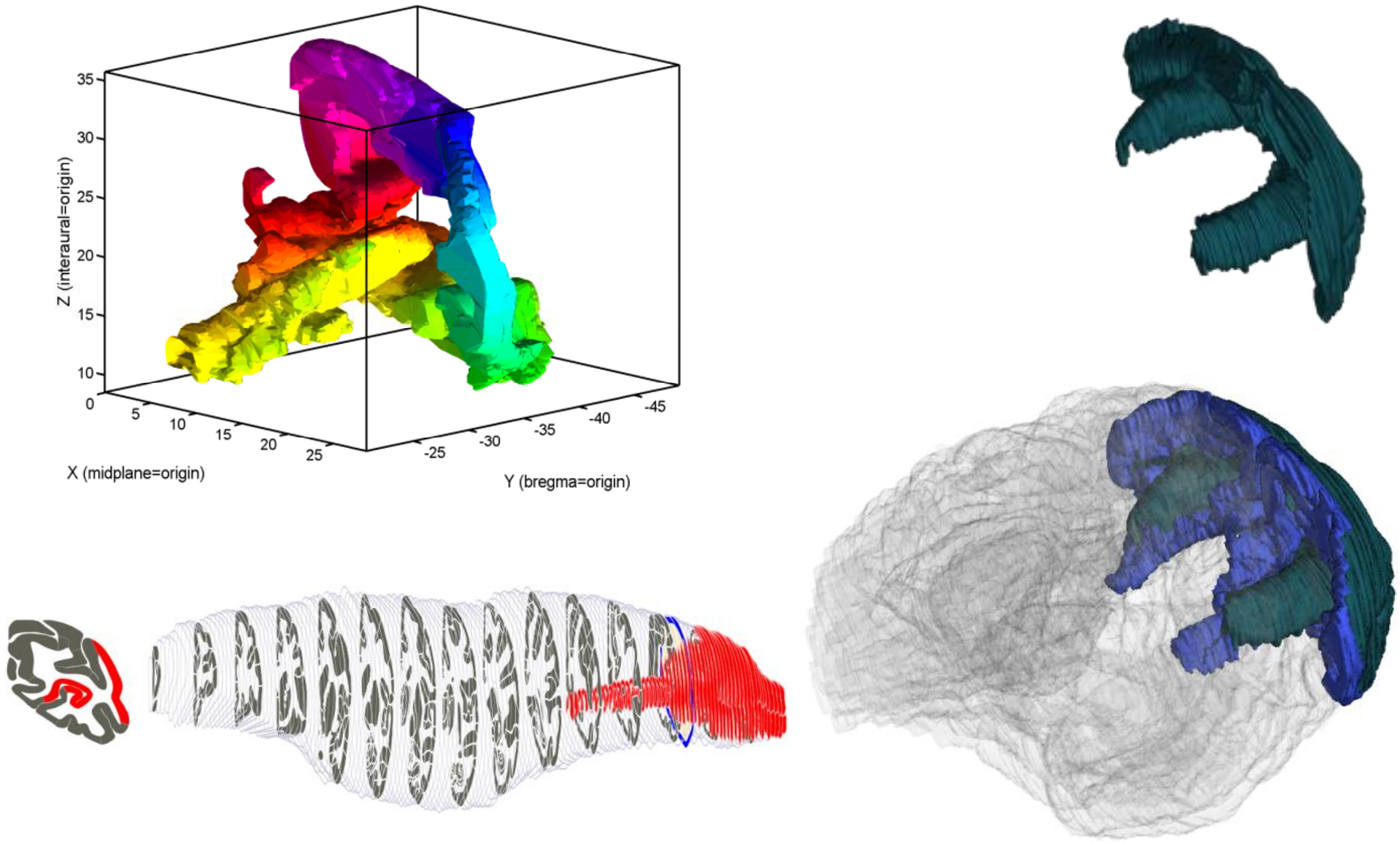

**Figure 4. Various ways to interact with brain region shapes, with PHT00-V2 as an example: (A) Using Matlab functions to download a template, extract a region mask and display it (scripts at http://scalablebrainatlas.incf.org/howto/analyze_templates_in_matlab.php); (B) Using the SBA thumbnail service; (C) Using the 3dBAR plugin from within SBA; (D) Using the 3dBAR custom reconstruction service (service.3dbar.org), showing both hemispheres, two areas (V1,V2) and a transparent whole brain.**

### 4.5 Label volume service (services/rgbslice.php)

For a given *template* and *slice number*, this service generates the SVG representing a coronal slice with color-coded brain regions, and converts it to a raster image if desired. By calling this service for each slice, a web-client can reconstruct the label volume. Matlab (The MathWorks Inc.) scripts are provided to download a complete atlas and visualize individual brain regions, as illustrated in Fig. 4A for template PHT00 and region V1.

### 4.6 Thumbnail service

For a given *template* and brain *region*, this service generates thumbnail images that can be used by other websites to illustrate what a brain region looks like and where in the brain it is located. Demand for this service came from the NeuroLex online semantic wiki for neuroscience terms (Larson et al. 2013). The service provides a choice of thumbnail layouts and image sizes. The output for the combined 2d and 3d view is illustrated in Fig. 4B.

## 5. Plugins

This section describes eight general purpose plugins, implemented in Javascript, often with an additional request handler service that resides on the SBA host (see Sec. 2.3). Some templates have additional plugins available, for example to see original data sets in high resolution.

### 5.1 Three-dimensional  Brain Atlas Reconstructor (3dBAR)

This plugin enables the user to view three-dimensional reconstructions of brain regions from 3dbar.org (Majka et al. 2013), as illustrated in Fig 4C for area PHT00-V1. To achieve this, data exchange routines were created to allow the SBA to import atlas templates from the 3dBAR-native CAF format, and 3dBAR to import templates directly from the SBA. The plugin shows precomputed thumbnails, and links to the 3dBAR service where the user can construct complex three-dimensional scenes, as illustrated in Fig. 4D.

### 5.2 Neuroscience lexicon (NeuroLex)

NeuroLex (Larson et al. 2013) is an online semantic wiki that aims to be the most complete and up to date reference work on neuroscience terms and concepts. It is a component of the NIF. For many brain regions it contains a definition and names of subregions, superregions etc. This plugin actively checks whether the selected brain region in the SBA has a representation in NeuroLex and if so, it downloads and presents the available properties, as illustrated in Fig. 5A for the Thalamus. NeuroLex in turn uses the SBA thumbnail service to graphically display the region.

### 5.3 BrainInfo and NeuroNames

http://www.braininfo.org is a web portal that contains detailed information on brain sites that are part of the NeuroNames ontology (Bowden et al. 2012). It contains data in the categories: Synonyms, Internal Structure, Cell types, Genes expressed, Locus in brain hierarchy, Connections, and Models. The plugin checks whether BrainInfo has a page about the currently selected brain region. BrainInfo does not currently have a service that returns structured data, and therefore the plugin is limited to displaying links to the corresponding page.

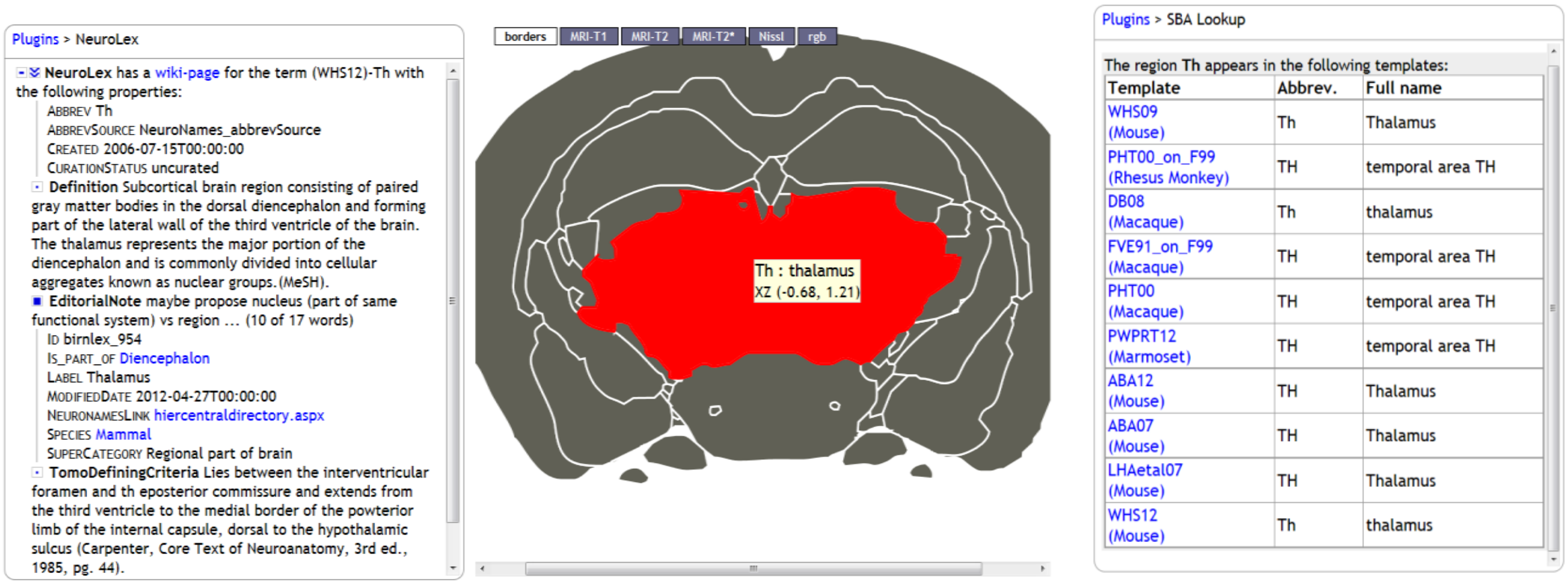

**Figure 5. Output of the NeuroLex (A) and SBA Lookup (B) plugins, both with Th (Thalamus) as the selected region.**

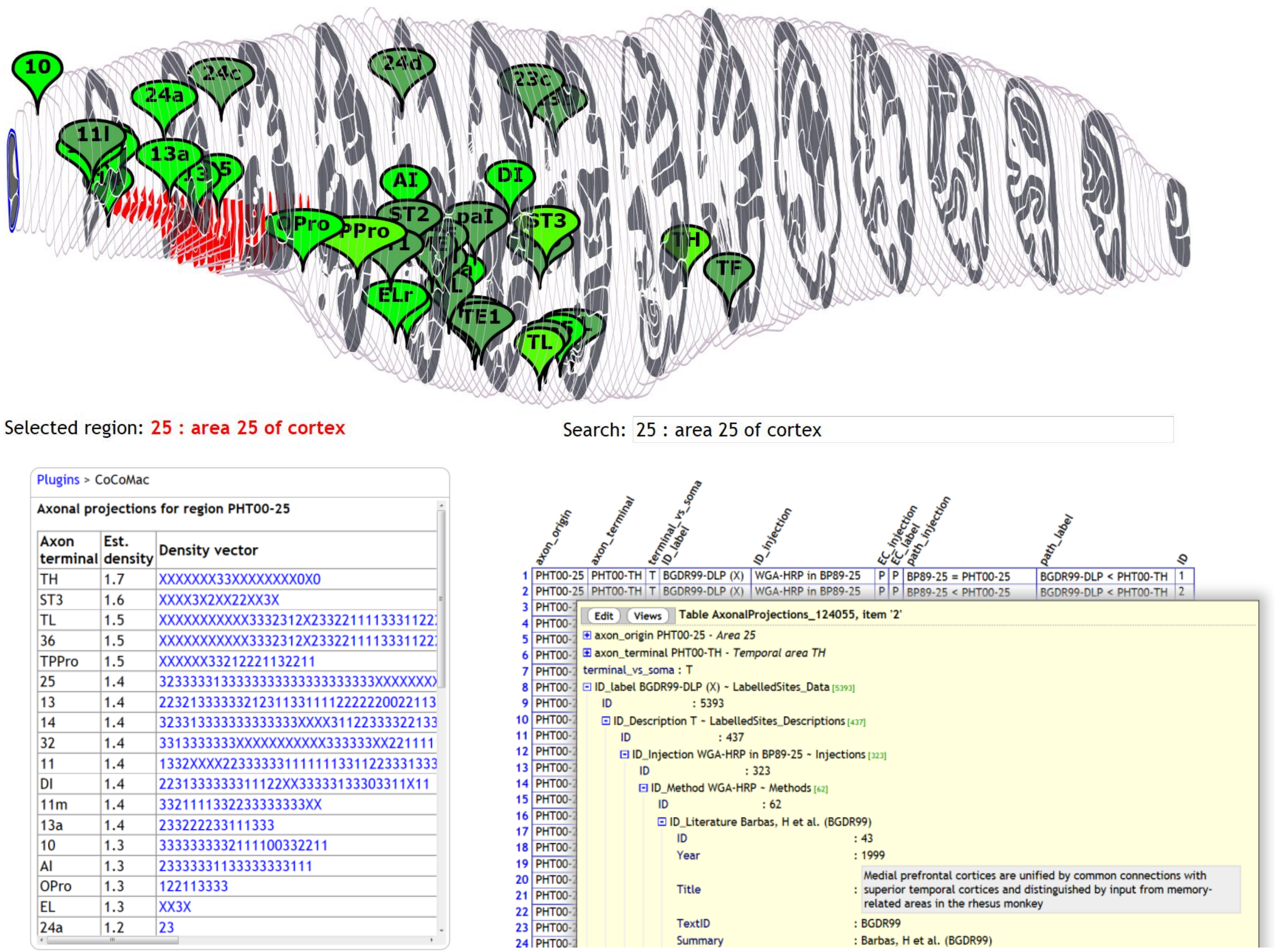

**Figure 6. Output of the CoCoMac plugin: (A) the axonal projections of region PHT00-25 are displayed as markers with a color intensity that represents connection strength. Note that this strength measure is not an official CoCoMac variable, it is provided to display capabilities of the SBA; (B) tabular output, in which each connection is represented by a character string. Each character is a separate 'piece of evidence', whereby X,0,1,2,3 mean unknown strength, absent, weak, medium and strong tracer labelling, respectively; (C) Interactive tabular display at the CoCoMac.g-node.org website allows traceback to the original publication.**

## 5.4 Wikipedia

Wikipedia (http://en.wikipedia.org) is a collaboratively edited, Internet encyclopedia that contains over 4 million articles in English. The plugin dynamically displays Wikipedia content that matches the full name of the currently selected brain region. Unlike NeuroLex, Wikipedia does not have attributes to limit results to neuroscience terms, and ambiguities with non-neuroscience terms may arise.

## 5.5 Stereotactic markers and transformations (AddMarker)

This plugin enables the placement of visible markers at a given stereotactic location, and displays the location of the last mouse click. For the WHS12 template, the plugin has the additional functionality of transforming coordinates to corresponding locations in other mouse atlas templates. The transformations are provided by the INCF Digital Atlasing Infrastructure (DAI, Hawrylycz et al. 2011). The result is illustrated in Fig. 3 where the center of the Interpeduncular nucleus in the WHS12 template is transformed to the Allen mouse reference atlas (ABA12).

## 5.6 Macaque connectivity (CoCoMac)

This plugin demonstrates the SBA at its full potential. It downloads Macaque structural connectivity data from the new CoCoMac database (Bakker et al. 2012) for the selected region, and displays each axonal projection as a marker positioned at the center of the connected region. Figure 6 shows the outgoing projections of region as a set of markers in the 3d panel, as a table in the plugin panel, and in detail on the CoCoMac.g-node.org website, where each projection can be tracked down to the publications in which they were reported.

### 5.7 Brain region lookup (SBA Lookup)

With the growing coverage of species and atlas templates, SBA is becoming a resource of its own. This plugin searches all templates for regions that have the same acronym, full name or alias as the currently selected structure, and provides direct links to the corresponding SBA pages. Figure 5B shows the importance of using the full region names to disambiguate the acronym-based results.

### 5.8 Fiducial points (Landmarks)

This WHS12-only plugin presents a set of 16 fiducial points that have been validated to be clearly recognizable on the basis of structural MR scans. Figure 7 displays the 3d panel with the landmarks. A workflow is under development to register new whole-brain volumes to the WHS12 reference space on the basis of (a subset of) these 16 landmarks (Sergejeva et al. 2014).

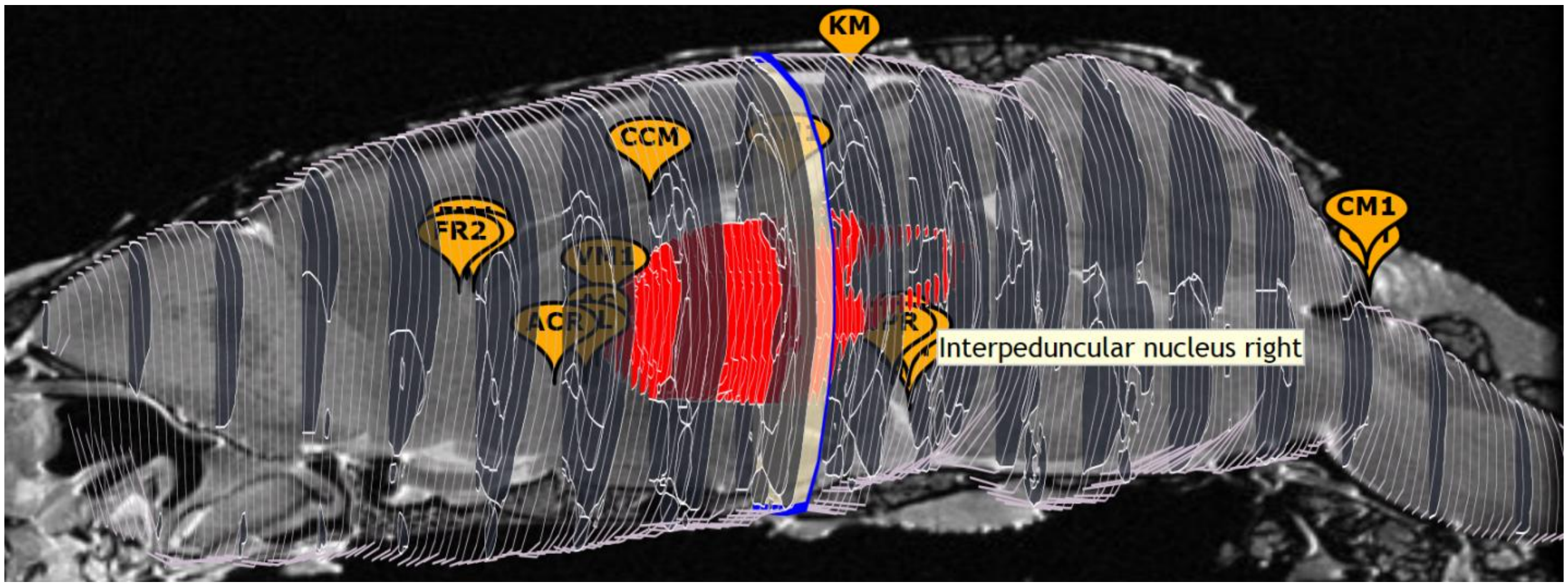

**Figure 7. Sixteen fiducial points, shown by the Landmarks plugin in the 3d panel of the WHS12 template, overlaid on the mid-saggital T1 slice. Semi-occluded markers are highlighted on mouse hover. Abbreviations: CM Cerebellum middle, KM Cortex middle, PM Pontine nucleus middle, HM Hippocampus middle, IP Interpeduncular nucleus middle, IPL/IPR Interpeduncular nucleus left/right, CCM Corpus Callosum middle, VM Ventricle middle, ACL/ACR Anterior Commissure left/right, FM Frontal middle, FL/FR Frontal left/right, AC Anterior Commissure.**

## 6. Discussion and Conclusion

What started as a simple CoCoMac visualization application based on manually redrawn region shapes, has grown into a comprehensive web toolkit that supports multiple species, multiple atlases, (third party) plugins, a self-search engine (SBA Lookup) and the ambition to expose all public atlasing resources that are of sufficiently high quality in a public, web-based interface. It attracts about 300 unique visitors per week. The SBA has made a first step towards the integration of data across templates and species with the ‘SBA Lookup’ plugin. The site is actively maintained, and four new services that will increase interaction and data integration are under way.

The first development is to support the display of saggital and axial slices in the 3d panel. A preview of this feature is show in Fig. 7, where the mid-saggital slice is combined with coronal SVG region contours.

The second development is a fully automated pipeline to import new atlas templates. The major hurdle was that the tools to vectorize multi-label images either produce poor quality results or need a manual curation step. This obstacle has recently been cleared with the development of the vectorization tool *mindthegap* (Kohli et al. 2014).

The third development will combine the automated atlas template pipeline with a nonlinear image registration step. This will superceed the current coordinate transformations as shown in Fig. 3. It will allow users who have volumetric data (MRI volume or Nissl stack) to view their data in conjunction with a (nonlinearly warped) region parcellation from one of the SBA templates.  The inverse scenario, whereby a user-provided volume is warped to fit in an existing SBA template, will also be supported. Harder to achieve is the registration of user-proveded single-slice data. A landmark-based workflow (Sec. 5.8) will allow rough positioning of the slice , but more accurate results require an image server that reslices brain volumes at arbitrary angles. While the technology to do so exists (Gustafson et al. 2007) this is beyond the current scope of SBA. For mouse and macaque, we recommend the NeuroMaps Mapper service (http://neuromaps.braininfo.org).

The fourth development is that SBA will be equipped to host atlases at a resolution of up to 2000 pixels in each dimension. At present, SBA does not store such data, but rather displays downsampled images with about 500x500 pixels in the non-coronal, and 180 pixels in the coronal dimension. High resolution data is only available through plugins that link to external resources. A *deep zoom* plugin will enable responsive display of high resolution content and will make the SBA suitable as a primary host for newly developed atlases.

An obvious omission from the SBA are several popular atlases that have previously appeared in print. There is no technical limitation to import such atlases, but the practice of transferring copyrights to the publishers prevents us from parsing such content. We try to convince copyright owners to become partners in the SBA project.

To conclude, it is our hope that this publication generates new initiatives for plugins, and we look forward to support inclusion of them in SBA. One idea for a community plugin is to have all regions in all supported templates mapped to a common ontology, such as the one developed by Puelles et al. (2013) or NeuroNames (Bowden et al. 2012). We will continue to develop our ‘flagship’ CoCoMac plugin with new levels of interactivity.

We invite owners of atlasing data to contribute and turn the Scalable Brain Atlas into a community driven resource.

## 7. Information Sharing Statement

All services of the Scalable Brain Atlas are accessible through the url http://scalablebrainatlas.incf.org. The source code for the Scalable Brain Atlas web services is available at https://github.com/INCF/Scalable-Brain-Atlas. The source data for each template can be downloaded as a set of JSON files described in Appendix A, license restrictions from the respective data owners do apply. Code related to importing new atlas templates is partly based on commercial software and is available on request. An open source release is in preparation, its 'mind-the-gap' vectorization engine is already available at https://github.com/INCF/Vectorization-of-brain-atlases.

## 8. Conflict of Interest

The authors have no conflict of interest.

## 9. Acknowledgements

The Scalable Brain Atlas is developed with joint financial support from the International Neuroinformatics Coordinating Facility (INCF) and the Donders Institute for Brain, Cognition and Behaviour of the Radboud University and UMC Nijmegen. The CoCoMac plugin is supported by the German INCF Node (BMBF grant 01GQ0801), Helmholtz Association HASB and portfolio theme SMHB. JUGENE Grant JINB33, and EU Grant 269921 (BrainScaleS). The work was conducted in the context of two INCF Programs: Ontologies of Neural Structures (PONS) and Digital Brain Atlasing (DAI). Inclusion of the Waxholm Space rat template was supported by the European Union Seventh Framework Programme (FP7/2007-2013) under grant agreement n° 604102 (HBP). Development of image registration services was supported by the Netherlands eSciene Center, grant 027.011.304.The following people contributed to the services, plugins and templates (see table 1 for abbreviations) that constitute the SBA: Daniel Wojcik and Piotr Majka (3dBAR plugin, whole brain 3d renderings, Marmoset template), Andreas Hess and Marina Sergejeva (Landmarks plugin), Hironobu Tokuno, Marcello Rosa and Tristan Chaplin (Marmoset template), Thomas Wachtler, Markus Diesmann (CoCoMac plugin), Jyl Boline, Janis Breeze (INCF taskforce integration), Doug Bowden (DB08 template, NeuroNames expertise), Gleb Bezgin (PHT00 and RM_on_F99 template and inspiration), Simon Eickhoff (EAZ05 template), Allan Johnson, Seth Ruffins (WHS template), Stephen Larson, Maryann Martone (bidirectional NeuroLex plugin), David van Essen (templates derived from Caret/SumsDB), Jan Sijbers and Jelle Veraart (VLAetal11 template), Henry Kennedy (citation policy).

# Appendix A: description of JSON files that make up an atlas template

see online document: http://scalablebrainatlas.incf.org/docs/template_data.php

## Appendix B: example plugin code

The code below shows the code to create a very basic plugin that prints the text “Hello World”, and a list of SBA state variables that the plugin can use to initiate queries. This code is available for download from the plugins directory of the SBA: http://scalablebrainatlas.incf.org/plugins/example_plugin.js.

All other plugins are also in that directory, under the name *pluginname_plugin.js*, where pluginname is the all-lowercase name of the plugin.

```
/*
 * Plugins are loaded after the global sbaViewer object has been
 * initialized, see ../js/sba_viewer.js for available methods of
 * the sbaViewer_class
 */

// Plugin constructor.
// Replace example by your own plugin name, in lowercase.
function examplePlugin_class(name,sbaViewer) {
  // JavaScript's way to call parent constructor
  sbaPlugin_class.apply(this,[name]);
  this.niceName = 'myExample';
  // store the atlas template name (e.g. PHT00) in this.template
  this.template = sbaViewer.template;
}
// JavaScript's way to create a derived class
examplePlugin_class.prototype = new sbaPlugin_class();

// Called when plugin is shown for the first time
examplePlugin_class.prototype.activate = function(sbaViewer,divElem) {
  var sbaState = sbaViewer.getState();
  divElem.innerHTML = 'HELLO WORLD<p/>'+json_encode(sbaState);
}

/*
// Called whenever the state of the SBA has changed
examplePlugin_class.prototype.applyStateChange =
function(sbaViewer,divElem) {
  // by default, this function calls the activate function,
  // you only need to define it when needed for improved efficiency.
}
*/
```